\documentclass[aps,prl,reprint,twocolumn,superscriptaddress,floatfix,nofootinbib,longbibliography]{revtex4-1}
\usepackage{amsthm}
\usepackage{amsmath,amssymb,color,comment,physics}
\usepackage[makeroom]{cancel}
\usepackage[caption=false]{subfig}
\usepackage{mathrsfs}
\usepackage{graphicx}
\usepackage{float}
\usepackage{subfig}
\usepackage[countmax]{subfloat}
\usepackage[english]{babel}
\usepackage{dsfont}
\usepackage[bookmarks=true,colorlinks,linkcolor=RoyalBlue,urlcolor=NavyBlue,citecolor=RoyalBlue]{hyperref}
\usepackage[dvipsnames]{xcolor}
\usepackage{braket}

\definecolor{mygold}{rgb}{0.93,0.59,0.13}
\definecolor{mypurple}{rgb}{0.49,0.18,0.56}

\newcommand{\ee}{\mathrm{e}}
\newcommand{\ii}{\mathrm{i}}
\renewcommand{\ket}[1]{\left|#1\right\rangle}
\renewcommand{\bra}[1]{\left\langle#1\right|}
\newcommand{\mean}[1]{\langle#1\rangle}

\usepackage{ulem}

\definecolor{mygreen}{rgb}{0.25,0.5,0.25}

\begin{document}

\preprint{APS/123-QED}

\title{Quantum scars and regular eigenstates in a chaotic spinor condensate}
\author{Bertrand Evrard}
\email{bevrard@phys.ethz.ch}
\affiliation{Institute for Quantum Electronics, ETH Z\"{u}rich, CH-8093 Z\"{u}rich, Switzerland}

\author{Andrea Pizzi}
\affiliation{Department of Physics, Harvard University, Cambridge, Massachusetts 02138, USA}

\author{Simeon I. Mistakidis}
\affiliation{ITAMP, Center for Astrophysics, Harvard $\&$ Smithsonian, Cambridge, Massachusetts 02138, USA}
\affiliation{Department of Physics, Harvard University, Cambridge, Massachusetts 02138, USA}

\author{Ceren B.~Dag}
\email{ceren.dag@cfa.harvard.edu}
\affiliation{ITAMP, Center for Astrophysics, Harvard $\&$ Smithsonian, Cambridge, Massachusetts 02138, USA}
\affiliation{Department of Physics, Harvard University, Cambridge, Massachusetts 02138, USA}

\begin{abstract}
Quantum many-body scars (QMBS) consist of a few low-entropy eigenstates in an otherwise chaotic many-body spectrum, and can weakly break ergodicity resulting in robust oscillatory dynamics. The notion of QMBS follows the original single-particle scars introduced within the context of quantum billiards, where scarring manifests in the form of a quantum eigenstate concentrating around an underlying classical unstable periodic orbit (UPO). A direct connection between these notions remains an outstanding problem. Here, we study a many-body spinor condensate that, owing to its collective interactions, is amenable to the diagnostics of scars. We characterize the system's rich dynamics, spectrum, and phase space, consisting of both regular and chaotic states. The former are low in entropy, violate the Eigenstate Thermalization Hypothesis (ETH), and can be traced back to integrable effective Hamiltonians, whereas most of the latter are scarred by the underlying semiclassical UPOs, while satisfying ETH. We outline an experimental proposal to probe our theory in trapped spin-1 Bose-Einstein condensates.
\end{abstract}

\maketitle

\textit{Introduction.}--- Isolated many-body systems out of equilibrium are typically expected to thermalize, meaning that the expectation value of generic physical observables reaches at long times the value predicted by a thermal ensemble. In this context, thermalization can be understood in terms of the Eigenstate Thermalization Hypothesis (ETH) \cite{PhysRevA.43.2046,1994PhRvE..50..888S,PhysRevLett.54.1879,Rigol2008,Rigolreview}, according to which most energy eigenstates are expected to be thermal \cite{PhysRevE.90.052105}. However, some systems escape this paradigm with mechanisms such as integrability \cite{kinoshita2006quantum,PhysRevLett.98.050405,PhysRevLett.103.100403},
Hilbert space fragmentation \cite{sala2020ergodicity,khemani2020localization}, many-body localization \cite{basko2006metal,schreiber2015observation,MBLreview1}
, and recently quantum many-body scars (QMBS) \cite{PhysRevLett.119.030601,bernien2017probing,Turner2018,moudgalya2018,2021NatPh..17..675S}. 
The latter leads to a weak ergodicity breaking phenomenon \cite{2021NatPh..17..675S}, whereby a few nonthermal states largely overlap with the initial condition. This results in robust oscillatory dynamics \cite{Turner2018}, as first experimentally observed in a Rydberg atom array \cite{bernien2017probing}.

The notion of quantum scars originates from quantum billards \cite{Heller2018,PhysRevLett.68.2867}. There, ``scars'' manifest in the eigenstate wavefunctions as a higher probability density in the vicinity of unstable periodic orbits (UPO) of the underlying chaotic classical motion \cite{PhysRevLett.53.1515}. Such an analysis cannot be directly performed on many-body quantum systems lacking a classical limit. Therein, QMBS are usually identified as nonthermal states embedded in the bulk of the energy spectrum \cite{2021NatPh..17..675S}. 
A connection to single-particle quantum scars was suggested for the PXP model in Ref.~\cite{PhysRevLett.122.040603} and further elaborated in Ref.~\cite{PhysRevX.10.011055}, relying on variational approaches to construct low-dimensional effective phase spaces. It was found that the periodic revivals associated to QMBS correspond to periodic orbits in the effective phase space. The region of phase space surrounding these periodic orbits is however regular (i.e., non-chaotic), unlike in scars. To date, the relation of QMBS to scars and regular states remains thus unclear~\cite{PhysRevX.10.011055}.

Here we address this problem from a fresh perspective, namely that of many-body systems with all-to-all interactions \cite{vardi2010bjj,mori2017classical,PhysRevA.97.023603,sinha2020chaos,rautenberg2020classical,2022arXiv221212046H,wittmann2022interacting}. While interacting, these systems have a well-defined classical limit, and thus allow to unambiguously discern regular and scar states. 
\begin{figure}[H]
\centering
\includegraphics[width=\linewidth]{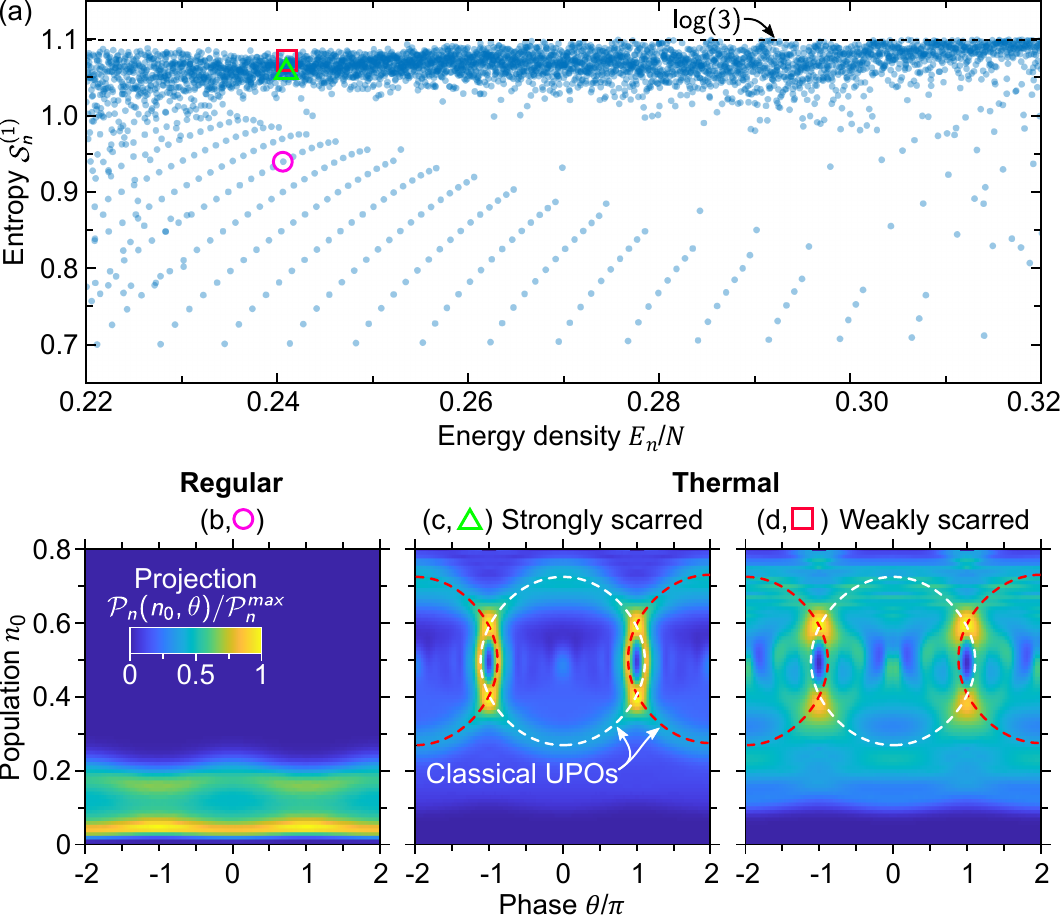}
\caption{\textbf{a} The entanglement entropy $\mathcal{S}^{(1)}_n$ of the eigenvalues of one-body reduced density matrix with respect to energy density $E_n/N$. $\mathcal{S}^{(1)}_n$ distinguishes thermal eigenstates, which nearly saturate the upper bound $\mathcal{S}^{(1)}_n=\ln3\approx1.1$, from an array of towers of regular eigenstates with lower entropy. \textbf{b-d} Phase space distributions $\mathcal{P}_n(n_0,\theta)$ of selected eigenstates, corresponding to the purple circle, green triangle and red square markers in panel (a). Regular states (b) lie within a restricted region of the phase space, whereas the scarred states (c,d) concentrate around the classical UPOs (white and red dashed lines) to some variable degree (strong for c, scarness $D_n\approx2.1$, and weak for d, $D_n\approx1.2$). Here, we considered $N=200$ particles.}\label{figure1}
\end{figure} 
More specifically, we introduce a chaotic model based on the spinor condensates \cite{kawaguchi2012spinor,stamper2013spinor} which are established experimental platforms to probe quantum many-body physics out-of-equilibrium \cite{Chang2005,Sadler2006,Pr_fer_2018,huh2023classifying}. We find that the majority of the eigenstates are thermal, featuring close-to-maximal entropy as expected for a chaotic system, see Figure~\ref{figure1}(a). Interestingly, however, most of these thermal states are scarred by a UPO in the underlying classical phase space [Figure~\ref{figure1}(c-d)], to an extent that we quantify via a scarness figure of merit. This ubiquity of scarring \cite{PhysRevLett.53.1515,Keski-Rahkonen_2019, PhysRevE.102.020101, Pilatowsky_Cameo_2021} stands in contrast to the usual phenomenology of QMBS. On the other hand, a smaller fraction of the Hilbert space consists of athermal eigenstates, which feature low entanglement entropy, violate ETH, and are associated to a regular region in the underlying phase space [Figure~\ref{figure1}(b)]. These regular states are organized in ``towers'' and can be approximately reproduced using a spectrum generating algebra. In light of these properties, they are reminiscent of QMBS. We conclude by investigating the dynamical signature of regular and scarred eigenstates: While the former give rise to robust oscillatory dynamics and prevent thermalization, the latter does not, but can cause a finite revival of the time-evolved state fidelity before thermalization. Our work sheds light on the fundamental differences between two notions of scars  found in the single- and many-body systems \cite{PhysRevLett.53.1515,2021NatPh..17..675S,regnault2022quantum,moudgalya2018}, and puts spinor condensates forward as a prominent platform for the experimental investigation of scarring. 

\textit{Model.}---
We consider $N$ spin$-1$ bosonic atoms tightly confined in an optical trap, such that spin and spatial degrees of freedom decouple, and condensation occurs in a single spatial orbital \cite{PhysRevLett.81.5257,PhysRevLett.81.742,yi2002single,PhysRevA.100.013622,PhysRevLett.126.063401}. In this limit, the energy scale associated with a spatial mode excitation is much larger than that of a spin excitation. The low energy physics is therefore governed by a spin Hamiltonian, reading
\begin{align}
\hat H_{\rm spin} = \frac{c_1}{N}\sum_{i< j}\hat{\bf s}_i\cdot\hat{\bf s}_j+\sum_i 
\left(
p_z\hat{s}_{z,i}+q\hat{s}_{z,i}^2+{\bf p}_\perp(t)\cdot\hat{\bf s}_{\perp,i}\,\label{eq.Hamiltonian1stQuant}
\right).
\end{align}
In Eq.~\eqref{eq.Hamiltonian1stQuant}, the first term describes an all-to-all Heisenberg interaction with strength $c_1/N$~\cite{kawaguchi2012spinor}. In the second sum, the first two terms correspond to the linear ($p_z\propto B_z$) and quadratic ($q\propto B_z^2$) Zeeman energies in the presence of a large static magnetic field $B_z$ along the $z$ axis. The last term is the linear Zeeman energy due to a sum of two weak fields rotating in the $(xy)$ plane at the frequencies $p_z\pm q$, thereby driving the hyperfine transition $0\leftrightarrow\pm1$. Performing a change of frame and a rotating-wave approximation, described in the Supplementary Material (SM)\,\cite{supp}, we obtain the following time-independent Hamiltonian
\begin{align}
		\hat H = \frac{c_1}{N}\left[\hat N_0(N-\hat N_0)+\frac{1}{2}(\hat N_+-\hat N_-)^2\right]+p(\hat W_++\hat W_-)\,, \label{eq.Hamiltonian2ndQuant}
\end{align}
where $\hat N_m=\hat a_m^\dagger\hat a_m$, with $\hat a_m$ the annihilation operator for the spin mode $m=0,\pm 1$, $\hat W_{\pm}=\frac{1}{\sqrt{2}}\left(\hat a_\pm^\dagger\hat a_0 + \textrm{H.c.}\right)$, and $p$ is the Larmor frequency associated to the rotating transverse field. An interesting feature of the interaction term in Eq.~\eqref{eq.Hamiltonian2ndQuant} is a logarithmic divergence of the density of states in the middle of the energy spectrum \cite{supp}. In contrast to previous studies of chaotic spinor condensates \cite{PhysRevLett.126.063401,rautenberg2020classical}, this key feature allows for the rapid emergence of chaos even for $p\ll c_1$. For concreteness, we will focus on $p=0.05$, near the onset of chaos, and set $c_1=\hbar=1$. 

To gain insights into the dynamics, we begin by deriving the corresponding semiclassical (mean-field) equations of motion. To this end, we define a SU(3) coherent state $\Ket{\zeta} = \frac{1}{\sqrt{N!}}[\sum_m \zeta_m\hat a^\dagger_m]^N\Ket{0}$
with $\zeta_m=\sqrt{n_m} e^{i\phi_m}$, where $n_m\equiv N_m/N$. Enforcing normalization, $\sum_m n_m = 1$, and choosing the global phase such that $\phi_0=0$, we can parametrize the coherent states by four real numbers, $n_0$, $\theta=\phi_++\phi_-$, $m=n_+-n_-$ and $\eta=\phi_+-\phi_-$ yielding mean-field equations of motion,
\begin{align}
\dot n_0= & \hspace{1mm} p\sqrt{2n_0}\bigg [\sqrt{n_+}\sin \phi_+ + \sqrt{n_-}\sin\phi_- \bigg ]\,,\notag\\
\dot \theta=& \hspace{1mm}2(1-2n_0) +p\bigg [\frac{2n_+-n_0}{\sqrt{2n_0n_+}}\cos\phi_+ \notag \\
&+\frac{2n_--n_0}{\sqrt{2n_0n_-}}\cos\phi_-\bigg ]\,,\notag\\
\dot m=& \hspace{1mm}p\sqrt{2n_0}\bigg [-\sqrt{n_+}\sin\phi_++\sqrt{n_-}\sin\phi_- \bigg]\,,\notag\\
\dot\eta=& \hspace{1mm}-2m-p\sqrt{\frac{n_0}{2}}\bigg [\frac{\cos\phi_+}{\sqrt{n_+}}-\frac{\cos\phi_-}{\sqrt{n_-}}\bigg ]\,,\label{eq:eqnMotion}
\end{align}
where $\phi_{\pm}$ and $n_{\pm}$ are functions of $n_0$, $m$, $\theta$, and $\eta$. 
The respective classical trajectories are obtained through numerical integration of these equations.

\textit{Quantum scars and regular eigenstates.}---
The presence of a clear classical limit in our model allows to visualize ergodicity breaking directly in phase space using the Husimi-Q distribution $Q_n(\zeta)=\left|\langle \zeta | \psi_n \rangle \right|^2$ of each quantum eigenstate $\ket{\psi_n}$, obtained through exact diagonalization of Eq.~\eqref{eq.Hamiltonian2ndQuant} for $N=200$ atoms. As it will be argued later, it is convenient for visualization purposes to focus on the $(n_0,\theta)$ plane, in which we define a projection function
\begin{align}
\mathcal{P}_n(n_0,\theta) = \frac{1}{d(n_0,\theta)}\iint dm d\eta \hspace{1mm} Q_n(n_0,\theta,m,\eta)\,,\label{eq.probadistrib}
\end{align}
where $d(n_0,\theta)=\iint dmd\eta \hspace{1mm} \delta(E_n-\bra{\zeta}\hat H\ket{\zeta})$ is the density of states at energy $E_n$ in the classical phase space.

\begin{figure}
		\centering
		\includegraphics[width=\columnwidth]{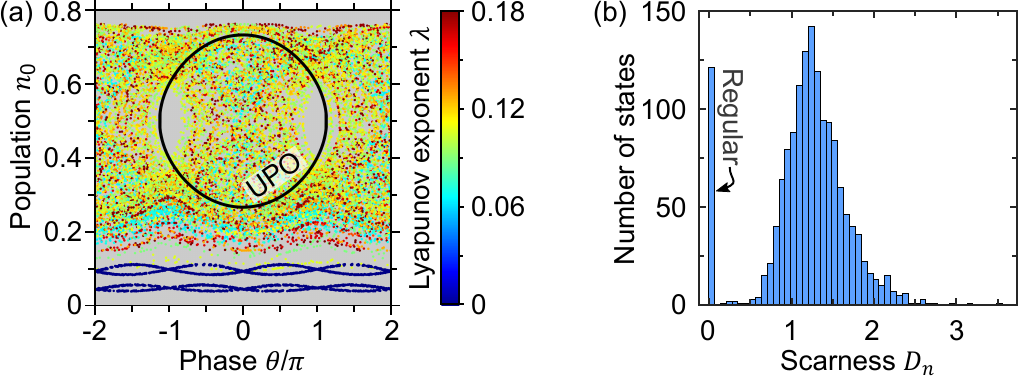}
		\caption{\textbf{a} Poincaré section of the classical equations of motion (\ref{eq:eqnMotion}) at energy $E/N=0.24$, showing a mixed phase space. The colors represent the Lyapunov exponent $\lambda$ of trajectories in phase space. The UPO at the same energy is represented by a solid black line ($\lambda_{\textrm{UPO}}=0.12$). The dark blue dots at $n_0 \ll 1$ correspond to stable periodic orbits ($\lambda\approx 0$). The density of states vanishes in the grey background. \textbf{b} Distribution of the scarness $D_n$ for $10^3$ eigenstates at energy $E=0.24N$.
  }\label{figure2}
\end{figure}

We present $\mathcal{P}_n$ for some characteristic eigenstates in Figure\,\ref{figure1}\textbf{b}-\textbf{d}. Remarkably, we find that most eigenstates exhibit a structure in phase space 
\cite{Keski-Rahkonen_2019, PhysRevE.102.020101, Pilatowsky_Cameo_2021}, even those (\textbf{c},\textbf{d}) with large entropy (see panel \textbf{a}). Indeed, the patterns observed in $\mathcal{P}_n(n_0,\theta)$ can be associated with periodic trajectories of the classical equations of motion, Eqs.~\eqref{eq:eqnMotion}. These lead to a mixed phase space with both regular and chaotic regions, as can be seen from an analysis based on Poincaré sections and Lyapunov exponents (Figure~\ref{figure2} \textbf{a}). Due to energy conservation, the motion is constrained to a three-dimensional manifold, and a 2D Poincaré section of equal-energy trajectories is thus obtained by fixing one variable. We choose to display $(n_0,\theta)$ for those times $t$ at which $m(t)=m(0)$. For each point on the Poincaré section, we extract the Lyapunov exponent $\lambda$ from the monodromy matrix \cite{supp}, and imprint it in the marker's color. For small $n_{0}$ we observe stable regular trajectories ($\lambda=0$). The motion becomes chaotic ($\lambda>0$) for larger $n_{0}$. Interestingly, it is seen from Eqs.~\eqref{eq:eqnMotion} that a trajectory starting on the 2D plane defined by $m=0$ and $\eta=0\,,2\pi$ remains on the plane. Evolving with Hamiltonian dynamics in 2D, these trajectories are periodic. As it turns out, however, some of these in-plane periodic orbits are unstable to out-of-plane perturbations, which is witnessed by a positive Lyapunov exponent $\lambda>0$.
These UPOs are responsible for scarring the eigenstates, see Figure\,\ref{figure1}\,\textbf{c}. 

We emphasize that scarring is very different from the correspondence between regular states and \textit{stable} (quasi)-periodic orbits. This can be understood from semi-classical arguments such as the Einstein–Brillouin–Keller quantization method for integrable models. This method applies when a \textit{volume} in phase space is regular, but cannot explain the scarring by UPOs which constitute a measure zero set \cite{berry1977regular,Heller2018}.
	
To quantify the scarring of the eigenstates $\ket{\psi_n}$, we define a ``scarness" figure of merit,
\begin{equation}
D_n = \frac{\oint_{\rm UPO} d\zeta \left | \bra{\psi_n} \zeta \right\rangle|^2}{\oint_{\rm UPO} d\zeta \left | \bra{\psi_e} \zeta \right\rangle|^2}.
\end{equation}
Here, the numerator (denominator) quantifies the overlap of the UPO with an eigenstate $\ket{\psi_n}$ (ergodic state $\ket{\psi_e}$) at energy $E_n$. The ergodic state $\ket{\psi_e}$ is built as a superposition of $10^4$ coherent states randomly picked along a classical chaotic trajectory at energy $E_n$. We thus expect $D_n\approx1$ if $\ket{\psi_n}$ is not scarred, $D_n>1$ if it is scarred by the considered UPO, and $D_n<1$ if $\ket{\psi_n}$ is scarred by another UPO or if it is a regular state. In Figure~\ref{figure2} \textbf{b}, we show the histogram of the values of $D_n$ in the middle of the energy spectrum. The peak at $D_n\approx 0$ is due to the regular states, which fill a region of phase space not explored by the UPO. Except this, the distribution of $D_n$ is biased towards values $D_n>1$ (excluding the regular states, the mean value of $D_n$ is $\approx 1.3$), suggesting that most eigenstates are scarred by the UPOs that we have previously identified. Note that alternative measures of scarness have been previously proposed \cite{2016NatSR...637656L,PhysRevLett.123.214101,Pilatowsky_Cameo_2021}. For rotating frequencies $\omega \pm q$, and $\omega \neq p_z$, a linear Zeeman term appears in Eq.~\eqref{eq.Hamiltonian2ndQuant} such that the UPO, and hence the scars, disappear.

QMBS feature low entropy and violate ETH, two key properties that we now investigate for our scarred and regular states. Unlike QMBS, we could not find an entropic measure that is significantly affected by the scarring. First, we inspect the von-Neumann entropy $\mathcal{S}^{(1)}_n=-\mathrm{Tr}(\rho^{(1)}_{n}\ln\rho^{(1)}_{n})$ of the one-body density matrix $[\rho^{(1)}_{n}]_{jk}=\bra{\psi_n}\hat a_j^\dagger\hat a_k\ket{\psi_n}$, which quantifies the entanglement between one atom and the rest of the ensemble \cite{evrard2021observation}. This entropy is large for all scarred states, c.f.,~Figure~\ref{figure1}, with no appreciable dependence on $D_n$. We have also checked the mode entanglement entropy \cite{vedral2003entanglement}, which is similarly large, while slightly anti-correlated with the degree of scarness $D_n$ (see SM \cite{supp}).

\begin{figure}
		\centering
		\includegraphics[width=\linewidth]{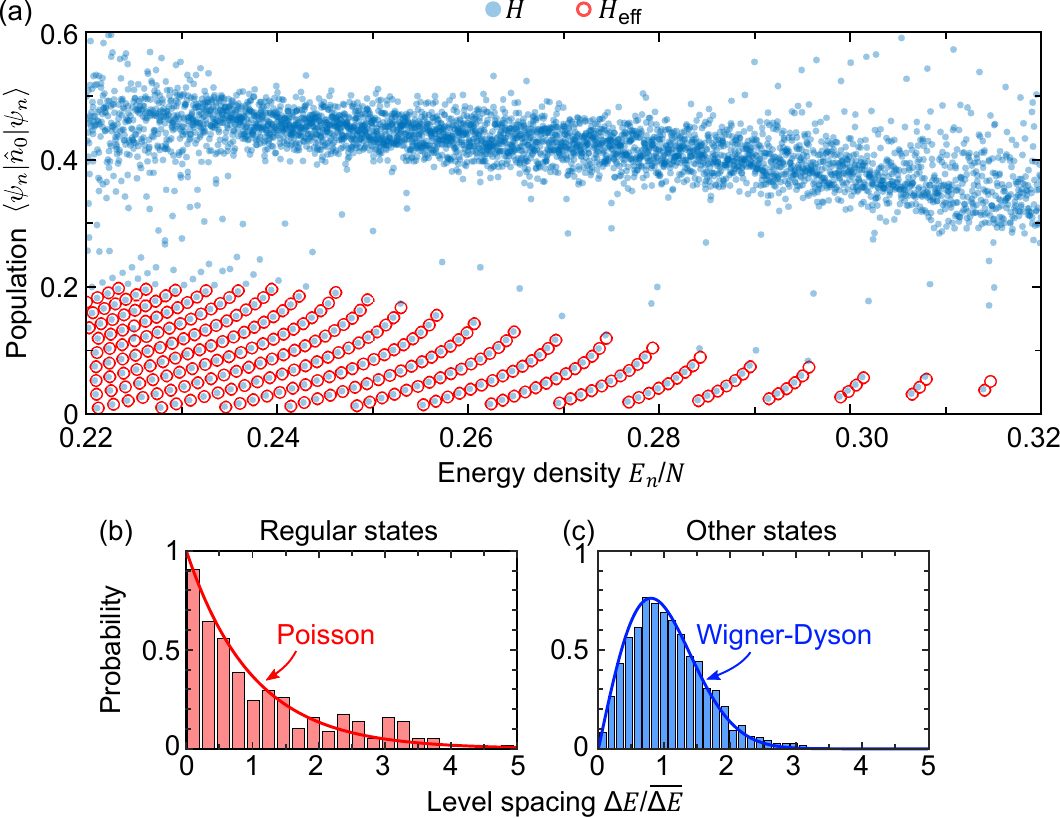}
		\caption{ \textbf{a} Eigenstate expectation values $\mean{\hat n_0}$, showing a narrow branch of thermal states with small fluctuations of $\mean{\hat n_0}$, in agreement with ETH, and towers of regular states violating ETH. The red 
 circles are obtained from a numerical diagonalization of the effective Hamiltonian (\ref{eq.Heff}). Level spacing statistics for the regular (\textbf{b}) and thermal (\textbf{c}) states. The solid lines depict the Poisson and the Wigner-Dyson distributions respectively.}\label{figure3}
\end{figure}

Let us now focus on the thermalization properties and energy level statistics. Figure\,\ref{figure3}\textbf{a} shows the expectation values of $\hat n_0=\hat N_0/N$ for each eigenstate $\ket{\psi_n}$, where $\Braket{\hat n_0}$ is a commonly measured quantity in spinor experiments \cite{PhysRevLett.123.113002,PhysRevA.100.013622,PhysRevLett.126.063401}. (similar behaviors are found for other few-body observables). Across the range $0.22<E/N<0.32$, there is a band of thermal states where $\mean{\hat n_0}$ is a smooth function of energy, one of the defining properties of ETH which can be rigorously demonstrated with the ETH indicators \cite{PhysRevA.97.023603} (see SM\,\cite{supp} for details). By contrast, $\mean{\hat n_0}$ is organized in towers for the regular states, having no clear dependence on energy, and thus violating ETH \cite{supp}. Consistently, in Figure\,\ref{figure3}\textbf{b}, \textbf{c} we demonstrate that the level spacing statistics is close to the Wigner-Dyson and Poisson distributions for thermal and regular states, respectively. 

These findings suggest that the regular states can be obtained from an underlying integrable model. Indeed, we are able to unravel it with the following reasoning: When there is a significant imbalance $N_+ \gg N_-$, we expect bosonic amplification of the mixing between $m=0$ and $m=+1$, leaving $m=-1$ as a background mode. In this case $\hat W_-$ can be neglected, the conservation of $\hat N_-$ is restored, and the eigenstates are Fock states for the $m=-1$ mode. We can thus replace the operator $\hat N_-$ by its eigenvalue $N_-$, and obtain the following integrable Hamiltonian for the remaining $m=0,+1$ modes
\begin{align}
\hat H_{\rm eff}(N_-) = -\frac{\hat N_0^2}{2N}+\frac{2 N_-}{N}\hat N_0+p\hat W_++\textrm{C} \,,\label{eq.Heff}
\end{align}
where $C$ is a constant \cite{supp}. For a fixed $N_-$ we diagonalize numerically $\hat H_{\rm eff}(N_-)$, and find eigenstate expectation values that match remarkably well those of a given tower. Considering a wide range of $N_-$ we are able to reconstruct all the towers seen in Figure\,\ref{figure3}\,\textbf{a}. Note that, to mitigate finite size effects, we diagonalize the effective Hamiltonian for $N=2000$ (instead of $N=200$ for the full Hamiltonian), and examine one eigenstate in every ten (details in SM \cite{supp}). There are $\propto N$ towers of $\propto N$ states, yielding a total number of regular states $\propto N^2$. As the size of the Hilbert space also scales as $\sim N^2$, the regular states constitute a finite fraction of it, even in the thermodynamic limit. Indeed, the regular states occupy a finite volume of the classical phase space, at low $n_0$, see Figure\,\ref{figure1}\textbf{b}.

In the limit $N_0\ll N_+$, Eq.~\eqref{eq.Heff} can be linearized and diagonalized via a Bogoliubov transformation. Hence the regular states can be constructed analytically from a set of two spectrum generating algebras, a common tool to construct QMBS \cite{2021NatPh..17..675S,moudgalya2018,Turner2018} (although these algebras here are approximate and only capture some regular states, see SM \cite{supp}). Within this approximation, the towers of regular eigenstates can be seen as a set of independent harmonic oscillators, explaining the Poisson energy level statistics. Finally, note that another set of regular states can be obtained upon inverting the role of the $m=\pm1$ modes, leading to nearly-degenerate doublets of regular states.

\begin{figure}
		\centering
		\includegraphics[width=\linewidth]{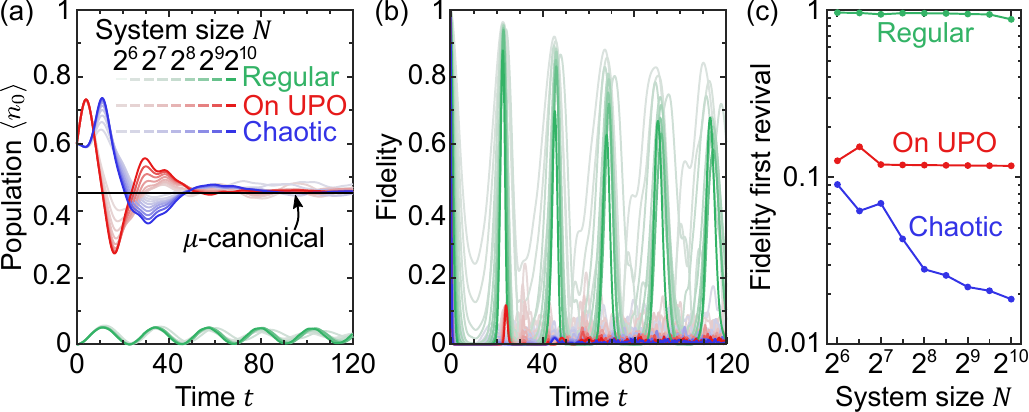}
		\caption{Time evolution of (\textbf{a}) $\mean{\hat n_0}$  and (\textbf{b}) the fidelity for an initial Fock state overlapping with regular states (green lines), a coherent state initiated on a UPO (red lines) and a coherent state in the chaotic sea (blue lines). The line intensity denotes increasing atom number (from $64$ to $N=1024$). In \textbf{a}, the solid black line is the microcanonical ensemble prediction with an energy window that excludes the regular states, and it is identical for all initial states which are at the same energy $E=0.24N$. In \textbf{c}, we show the fidelity achieved at the first revival as a function of the atom number.}\label{figure4}
\end{figure}
		
\textit{Dynamics.}---We now investigate the dynamics of the system to probe scarring, see Figure~\ref{figure4}. We consider three initial states at the energy $E=0.24N$: (i) a coherent state $\ket{\zeta_\mathrm{c}}$ with $n_0=(1+\sqrt{1-4E/N})/2$ and $m=\theta=\eta=0$, such that $\zeta_\mathrm{c}$ overlaps with the chaotic region of the phase space; (ii) a coherent state $\ket{\zeta_\mathrm{s}}$ with $n_0=(1+\sqrt{1-4E/N})/2$, $m=\theta=0$, and $\eta=\pi$, such that $\zeta_\mathrm{s}$ lies on the UPO; and (iii) a Fock-state $\ket{\psi_\mathrm{reg}}$  with $N_0=0$ and $N_+-N_-=\sqrt{2NE}$, chosen to have a large overlap with one tower of regular states. We focus on the observable $\mean{\hat n_0}$, though we observe similar behavior for other few-body observables. 

Regular states display long-lived oscillations in the time evolution of both $\mean{\hat{n}_0}$ and the state fidelity $\mathcal{F}(t)=|\bra{\psi(0)}\psi(t)\rangle|^2$, independently of the system size, see Figure~\ref{figure4}(a) and (b), respectively. On the other hand, when the system is initiated at the two coherent states $\ket{\zeta_\mathrm{c}}$ and $\ket{\zeta_\mathrm{s}}$ \cite{evrard2021}, which overlap with eigenstates satisfying ETH, $\hat{n}_0(t)$ rapidly thermalizes to the microcanonical ensemble prediction (excluding the regular states). The role of scarring becomes evident at the level of the fidelity $\mathcal{F}(t)$, Figure~\ref{figure4}(b): its first revival is stable in the $N\gg1$ limit for the initial state $\ket{\zeta_\mathrm{s}}$, whereas it vanishes for the initial state $\ket{\zeta_\mathrm{c}}$. This can be better appreciated from the scaling analysis in Figure~\ref{figure4}(c).

\textit{Discussion and Outlook.}---
We have studied the statistical and dynamical properties of a chaotic spinor condensate, in which a semiclassical limit allows to unambiguously discern coexisting quantum scars and regular eigenstates. We recover the properties of chaotic many-body systems with ETH-obeying and high entropy eigenstates, and Wigner-Dyson energy spacing distribution. 
Remarkably, these features arise despite the fact that most eigenstates are scarred \cite{Pilatowsky_Cameo_2021}, instead of random as expected for a chaotic system \cite{Rigolreview}. These results also highlight the difference between quantum scars  \cite{PhysRevLett.53.1515} and QMBS, which are often defined as athermal eigenstates \cite{2021NatPh..17..675S,regnault2022quantum,moudgalya2018} - a definition that in our system would point towards regular states. An interesting open question remains regarding the fate of regular and scar states in the presence of perturbations that would break the all-to-all nature of the model, e.g., when loading the spinor condensate in an optical lattice.

Our model's main ingredients are readily available in state-of-the-art experimental setups: a tight laser trap to freeze the spatial degrees of freedom, a large bias magnetic field to suppress spin-changing collisions, and a rotating transverse magnetic field to introduce mode-mixing. The required extremely stable magnetic fields could be achieved via magnetic shielding \cite{farolfi2019design}.
This is an exciting prospect given the scarcity of experimental observations of quantum scars \cite{PhysRevLett.68.2867,1996Natur.380..608W}, and it could provide an ideal testbed to investigate the relation between scarring and decoherence in a physical system with a weak coupling to the environment.

We thank Eric~J.~Heller, Joonas Keski-Rahkonen, Francisco~L.~Machado, Hannes Pichler, and Saúl Pilatowsky-Cameo for helpful comments and discussions. B.E. is supported by an ETH Zurich Postdoctoral Fellowship. A.P. is supported by the AFOSR MURI program (Grant No. FA9550-21-1-0069). C.B.D. and S.I.M. acknowledge financial support from the NSF through a grant for ITAMP (Award No: 2116679) at Harvard University.


\bibliographystyle{apsrev4-1}

%


\onecolumngrid
\newpage

\setcounter{equation}{0}
\setcounter{figure}{0}
\setcounter{table}{0}
\setcounter{page}{1}
\makeatletter
\renewcommand{\theequation}{S\arabic{equation}}
\renewcommand{\thefigure}{S\arabic{figure}}
\renewcommand{\bibnumfmt}[1]{[S#1]}

\begin{center}
\textbf{\Large Supplementary Material: Quantum scars and regular eigenstates in a chaotic spinor condensate}
\end{center}
\hspace{5mm}
\begin{center}
{\large Bertrand Evrard, Andrea Pizzi, Simeon I. Mistakidis and Ceren B. Dag }
\end{center}

\vspace{5mm}

This Supplementary Material is devoted to complementary analytical and numerical calculations in support of the the findings in the main text. Specifically, in the first section we discuss the experimental proposal  of our model utilizing a spinor Bose-Einstein condensate, and in particular examine how the regular eigenstates and scarring can be dynamically probed. In the second section, we perform a detailed numerical analysis to study the validity of the Eigenstate Thermalization Hypothesis (ETH) on the model under consideration.  We confirm that ETH holds for the chaotic band of states while it is violated for the regular eigenstates. In the third section, we present the effective description of the regular states in detail. In particular, we demonstrate the self-consistency of the approach, and how two (approximate) spectrum generating algebras can be derived from the effective Hamiltonian. Next, we briefly review the dynamical system formalism for calculating the Lyapunov exponents based on the fundamental matrix approach. In the last two sections, we derive the density of states for the interaction Hamiltonian, and study the behavior of the mode entanglement entropy.

\section{Proposal for Experimental realization}
\subsection{Implementation of the spinor Hamiltonian}

Our results can be realized using a spinor BEC \cite{kawaguchi2012spinor,stamper2013spinor} of alkali atoms, e.g. Sodium, being tightly confined in a harmonic trap. In their electronic ground state, alkali atoms have an hyperfine spin $F=1$. We consider the particular situation where the atoms are condensed in the same spatial mode \cite{yi2002single}, and the only relevant degrees of freedom are their Zeeman state $m=\pm1,0$, where $m$ is the quantum number associated to the projection of the hyperfine spin along a quantization axis --- chosen to be the axis of a bias magnetic field (see below). 
This is called the single-mode approximation (SMA)~\cite{stamper2013spinor,kawaguchi2012spinor} and it is valid for small atom numbers (typically, $N\sim10^2$ to $10^4$) when the condensate is prepared in a tight optical trap \cite{PhysRevLett.126.063401}. 
This approach can break down for long evolution times either due to the build-up of spatial correlations~\cite{stamper2013spinor,mittal2020many} as well as atom loss related processes from the condensate \cite{PhysRevA.100.013622,bookjans2011quantum,stamper2013spinor}. 
In the presence of a fixed magnetic field ${\bf B}=B{\bf e}_z$ of small magnitude, the spinor Hamiltonian accounting for the Zeeman effects reads 
\begin{align}
    \hat H_Z \approx p_z \hat S_z+q(\hat N_++\hat N_-) = p_z \hat S_z-q\hat N_0+\mathrm{const.} \,,    
\end{align}
where $p_z\propto B$ and $q\propto B^2$ are the linear and quadratic Zeeman shifts, and the expansion is valid for $q\ll p_z$. 
Also, the operator $\hat S_z=\hat N_+-\hat N_-$ represents the component of the total spin $\hat{\bf S}$ along the quantization axis $z$. Moreover, the spin-dependent interactions are described by 
\begin{eqnarray}
 \hat H_{S} = \frac{c_1}{2N}\hat{\bf S}^2\ =  \frac{c_1}{2N}\bigg [ 2\hat N_0(\hat N_++\hat N_-)+\hat N_+^2 - 2\hat N_+ \hat N_- +\hat N_-^2 +2(\hat a_0^2\hat a_+^\dagger\hat a_-^\dagger+\hat a_0^{\dagger2}\hat a_+\hat a_-)+2N-\hat N_+-\hat N_- \bigg] \,.\label{eq.total spin}
\end{eqnarray}
Here, the linear term can be absorbed in the quadratic Zeeman shift, and it will be thus omitted in the following. 
The operator $\hat{a}_0$ ($\hat{a}_{\pm}$) annihilate an atom in the $m=0$ ($m=\pm 1$) hyperfine state. 

Since $[\hat H_{S},\hat S_z]=0$, the quadratic Zeeman effect can not be neglected even in the $q\ll p_z$ regime. 
In order to engineer the Hamiltonian discussed in the main text, we need to include  linear mode mixing terms. This can be achieved by further introducing two rotating fields at frequencies $\omega_{\pm}=p_z\pm q$, being resonant with the transition $0\leftrightarrow\pm1$, namely
\begin{align}
    \hat H_{R}= &\frac{p}{2}\left[\cos(\omega_+t)+\cos(\omega_-t)\right]\hat S_x +\frac{p}{2}\left[\sin(\omega_+t)+\sin(\omega_-t)\right]\hat S_y\,.\label{eq: Hamiltonian modulation}
\end{align}
Let us now perform the unitary operation $\hat U = \exp(i\hat H_Zt)$ to obtain an effective Hamiltonian $\hat H'=\hat H_{S}'+\hat H_R'$ where 
\begin{align}
    \hat H_{S}'=\hat H_{\rm int}+\frac{c_1}{N}\left(e^{2iqt}\hat a_+^\dagger\hat a_-^\dagger \hat a_0^2+e^{-2iqt}\hat a_0^{\dagger2}\hat a_+\hat a_-\right)\,.\label{eq.Hint(t)}
\end{align}
In this expression, $\hat H_{\rm int}$ quartic term of the Hamiltonian given in the main text, and
\begin{align}\label{eq.Hrot(t)}
    \hat H_R' &= p\hat S_x+\frac{p}{\sqrt{2}}\left[e^{2iqt}(\hat a_0^\dagger\hat a_++\hat a_-^\dagger\hat a_0)+\textrm{H.c}\right] 
    = \frac{p}{\sqrt{2}}\left[(1+e^{2iqt})(\hat a_0^\dagger\hat a_++\hat a_-^\dagger\hat a_0)+\textrm{H.c}\right].
\end{align}
In the limit $q\gg |c_1|$ and $q \gg p$ we average out the oscillating terms in Eqs.\,(\ref{eq.Hint(t)}), (\ref{eq.Hrot(t)}) and obtain the target Hamiltonian. 
The main experimental challenge is to substantially reduce the magnetic field fluctuations, 
i.e. render them much lower than the spin-dependent interaction strength.  
For instance, in the case of a Sodium BEC with $c_1 \sim h\times30\,$Hz, the fluctuations should be of the order of $\mu G$, which has been demonstrated using magnetic shielding \cite{farolfi2019design}.

\subsection{Investigating the dynamics}

We now discuss how to experimentally monitor the dynamical behavior of the spinor BEC described  
in the main text. 
Our system is prepared in the polar state where all atoms lie in the $m=0$ mode, which can be prepared with very high fidelity using a magnetic gradient to pull out atoms from the $m=\pm 1$ states. 
Afterwards, different coherent states can be generated using a combination of rotating magnetic field pulses together with free evolution in a large magnetic field, see also the discussion in Ref. \cite{evrard2021}). 
First we consider a rotating magnetic field pulsed for a time $t_1$ in order to achieve a rotation in spin space. 
In the rotating frame, the operator expressed in the single-particle Zeeman state basis ($m=\pm1,0$) reads
\begin{equation}
    \exp\left(-\ii\alpha\hat s_x\right)=\begin{pmatrix}
        \frac{\cos\alpha+1}{2}&-\frac{i\sin\alpha}{\sqrt{2}}&\frac{\cos\alpha-1}{2}\\
        -\frac{i\sin\alpha}{\sqrt{2}}&\cos\alpha&-\frac{i\sin\alpha}{\sqrt{2}}\\
        \frac{\cos\alpha-1}{2}&-\frac{i\sin\alpha}{\sqrt{2}}&\frac{\cos\alpha+1}{2}
    \end{pmatrix}\,,
\end{equation}
where $\alpha = \Omega t_1$ and $\Omega$ denotes the Rabi frequency. 
After this first pulse, the mode populations correspond to $n_0=\cos^2\alpha$ and $n_+=n_-=\sin^2\alpha/2$. 
To complete the preparation, one needs to tune the phases $\phi_{\pm 1}$, which can be simply achieved via the linear and quadratic Zeeman shifts of the static field, which results in the evolution $\phi_\pm = (q\pm p_z)t_2$ during a hold time $t_2$. The parameters characterizing the coherent states used in the main text thus evolve as $\theta = 2qt_2$ as well $\eta = 2p_zt_2$, and by adjusting the time delay $t_2$ it is possible to obtain the desired initial state. Consequently, the magnetic field modulation as described by Eq.~(\ref{eq: Hamiltonian modulation}) is turned on initiating the dynamics. 
Note that in the above-described preparation steps we have neglected the effect of the interaction, which is valid when $c_1t_1,c_1t_2\ll1$, and can be easily achieved in practice. 
Within this scheme it holds that $n_+=n_-$, which is desirable condition to probe quantum scars. However, it is also possible to start from a state with all atoms in the $m=1$ hyperfine level and generate other coherent states aiming to probe the regular states.

Next, let us turn to the diagnostics of the dynamics. Using a Stern-Gerlach scheme, the populations $n_{0,\pm}$ can be measured simultaneously, and thus $S_z=N_+-N_-$ can be deduced. Rotation in spin space can be achieved utilizing pulses of the rotating magnetic field, giving access to any component of $\hat{\bf S}$. 
In this way, it is possible to observe the thermalization (or lack thereof) of one-body observables, but also even reconstruct the one-body density matrix \cite{evrard2021observation} and estimate the associated entropy growth. 
Measuring the fidelity between the time-dependent state and the initial one is certainly more challenging, but it can be achieved in principle for a coherent state. Indeed, let us consider the initial state
\begin{align}
\ket{\psi_i}=\ee^{-i\hat H_Zt_2}\ee^{-i\alpha\hat S_x}\ket{0}^{\otimes N}\,.
\end{align} 
To evaluate the time-dependent fidelity, under the Hamiltonian $\hat H$, we ``undo" the single-particle manipulation that was done in the above-described preparation steps, such that we obtain the state
\begin{align}
    \ket{\psi_f}=\ee^{-i\alpha'\hat S_x}\ee^{-i\hat H_Zt_2'}\ee^{-i\hat Ht}\ket{\psi_i}\,,
\end{align} 
where $\alpha'$ and $t_2'$ are chosen such that $\alpha'=-\alpha+n_12\pi$, $pt_2'=-pt_2+n_22\pi$ and $qt_2'=-qt_2+n_32\pi$, where $n_{1,2,3}$ are integers. Then, one measures 
the probability that all atoms are found in the $m=0$ state, which can be written as
\begin{align}
    \vert\bra{0}^{\otimes N}\ket{\psi_f}\vert^2=\vert\bra{\psi_i}\ee^{-i\hat Ht}\ket{\psi_i}\vert^2\,,
\end{align}
which is exactly the fidelity between the time-evolved state $\ket{\psi(t)}=\ee^{-i\hat Ht}\ket{\psi_i}$ and the initial one. 

\section{Eigenstate Thermalization Hypothesis properties of the model}

Here, we are aiming to numerically prove the Eigenstate Thermalization Hypothesis (ETH) for the chaotic band of states which are mostly scarred by a UPO as highlighted in the main text. 
Afterwards, it is shown that ETH is violated for the regular states.  

The ETH criteria for the observable $\mean{\hat{n}_0}$, i.e. the population density in the zeroth hyperfine level, with its eigenstate expectation values (EEV) defined as $n_{0,pq}=\Bra{\psi_p}\hat{n}_0\Ket{\psi_q}$ are the following \cite{Rigolreview},\\
(1) The diagonal elements, $n_{0,pp}$,  
should behave smoothly with respect to the energy eigenvalues $E_{p}$ \cite{PhysRevLett.54.1879,PhysRevA.43.2046,Rigol2008}. \\
(2) The off-diagonal elements, $n_{0,pq}$, should be negligibly small as compared to the diagonal elements $n_{0,pp}$, while $n_{0,pq}$ should be a smooth function of the energy differences $E_p-E_q$ \cite{MarkSrednicki_1996,MarkSrednicki_1999}.

Both statements could be formulated as
\begin{eqnarray}
    n_{0,pq} = n_0(E_p) \delta_{pq} + \text{max} |n_{0,pq}| f_{n0} (\bar{E},E_p-E_q) R_{pq},
\end{eqnarray}
where $\bar{E}=(E_p+E_q)/2$ is the mean energy, $R_{pq}$ is a random real or complex variable that satisfies $\overline{|R_{pq}|^2}=1$, while $n_0(E_p)$ and $ f_{n0} (\bar{E},E_p-E_q)$ are smooth functions in terms of their arguments. If this mathematical statement is valid within a narrow energy window $E_k \in [E_c-\delta,E_c+\delta]$ where the interval $\delta E \equiv 2\delta \ll E_c$, then the observable $\mean{\hat{n}_0}$ relaxes to its thermal value dictated by the microcanonical ensemble 
\begin{eqnarray}
    \rho_{\text{mc}} = \frac{1}{N_{\text{int}}} \sum_k \Ket{\psi_k}\Bra{\psi_k}.
\end{eqnarray}
This implies that any eigenstate in this energy window can predict the equilibration value of the observable at long evolution times, and hence it is thermal.

\begin{figure*}
\includegraphics[width=0.8\textwidth]{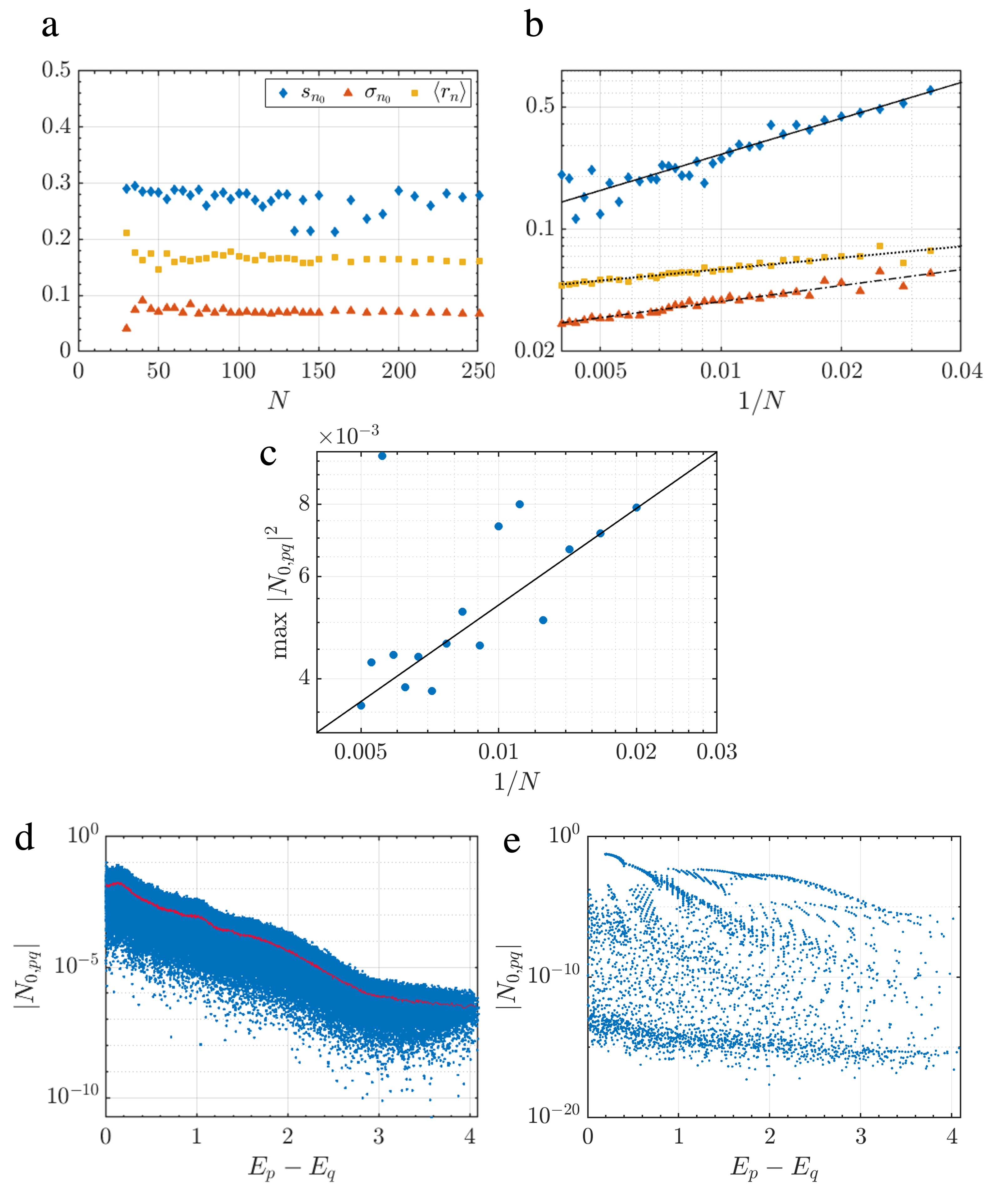}\caption{Indicators of Eigenstate thermalization hypothesis (ETH). \textbf{a}-\textbf{b} The first ETH criterion (diagonal ETH) is tested with the indicators of support $s_{n_0}$, noise $\sigma_{n_0}$ and eigenstate expectation value (EEV) nearest neighbor differences, $\langle r_{n_0} \rangle$. 
It is violated for regular states as depicted in panel \textbf{a} and it is satisfied for the chaotic states shown in panel \textbf{b}. \textbf{c}-\textbf{e} The second criteria of ETH (off-diagonal ETH) is checked, and found to be satisfied for chaotic states, see panel \textbf{c}, by noticing that the maximum off-diagonal element of the observable $N_0$ in a narrow fixed energy window decreases with increasing atom number $N$. \textbf{d} The off-diagonal elements, $N_{0,pq}$, 
behave smoothly with respect to the energy differences $E_p-E_q$ for chaotic scarred states where the red line refer to the moving time average. \textbf{e} $N_{0,pq}$ is not a smooth function of $E_p-E_q$ for regular states. Rather they are very scattered. As such, the second criteria of ETH fail for the regular states. }\label{SMFigure3}
\end{figure*}

We first examine the criteria for both the chaotic band of states, which contains quantum scars, and the regular states. It was already shown in Figure~3 of the main text that the diagonal elements of the EEV were smoother in the chaotic band of states as compared to the towers of regular states. Here, we make this observation quantitative by examining three ETH indicators: (i) Support $s_{n_0}$ in a fixed energy window, (ii) ETH noise $\sigma_{n_0}$ and (iii) nearest neighbor differences in the EEV, $\Braket{r_{n_0}}$, in an energy window which is a function of atom number, i.e.,~the energy window we choose increases proportionally with the atom number as we increase the atom number. These indicators are defined as follows
\begin{eqnarray}
    s_{n_0} &=& \max\left(\Bra{\psi}n_0\Ket{\psi}\right)_{\psi \in \delta E} - \min\left(\Bra{\psi}n_0\Ket{\psi}\right)_{\psi \in \delta E},\\
    \sigma_{n_0} &=& \frac{1}{\sqrt{N_{\text{int}}}}\sqrt{\sum_{\psi_n \in \delta E} \left[ \Bra{\psi_n}n_0 \Ket{\psi_n} - \Braket{n_0}_{\text{mc},\delta E}\right]^2 },\\
    \Braket{r_{n_0}} &=& \frac{1}{N_{\text{int}}} \sum_{\psi_n \in \delta E} \bigg|\Bra{\psi_{n+1}}n_0 \Ket{\psi_{n+1}} - \Bra{\psi_n} n_0 \Ket{\psi_n}\bigg|.
\end{eqnarray}
In these expressions, $\Braket{n_0}_{\text{mc},\delta E}$ denotes the microcanonical ensemble prediction of the observable $\hat n_0$, $\sum_{\psi_n\in \delta E}$ indicates summation over $N_{\text{int}}$ number of eigenstates that remain in the narrow energy window, and finally $f(\dots)_{\psi_n \in \delta E}$ indicates that the generic function $f$ is calculated over the eigenstates that remain in the window. Since $s_{n_0}$ is sensitive to outlier states, we fix the energy window for all atom numbers while computing this particular ETH indicator. 

We find that all these ETH indicators feature a decreasing trend for larger atom number $N$, see Figure~\ref{SMFigure3}b, concerning the chaotic band of eigenstates which have various degrees of scarness as explained in the main text. In sharp contrast, the ETH indicators are almost insensitive to atom number variations for the regular eigenstates, see  Figure~\ref{SMFigure3}a. 
Notice that $s_{n_0}$ estimates whether the EEV converges to a a fixed value in the thermodynamic limit. Figure~\ref{SMFigure3}b shows that the chaotic band indeed gets flatter as $N$ increases, whereas the regular towers in Figure~\ref{SMFigure3}a continue to span the same amount of space in the EEV. Specifically the scaling for the chaotic band follows as, $s_{n_0} \propto N^{-0.68}$. ETH noise quantifies the fluctuations of the EEV around the microcanonical ensemble prediction. We also observe that $\sigma_{n_0}$ decreases with the atom number obeying $\sigma_{n_0}\propto N^{-0.31}$ for the chaotic band, and this is consistent with vanishing support. Finally, $\Braket{r_{n_0}}$ averages over the nearest neighbor differences between the EEV. For an EEV distribution that tends to a uniform distribution, $\Braket{r_{n_0}}$ has to decrease with $N$. We observe $\Braket{r_{n_0}} \propto N^{-0.22}$. Let us note that the scaling exponents are bigger if the energy window for $\sigma_{n_0}$ and $r_{n_0}$ are fixed. The observation that all decay exponents are less than $2$ signifies that the first ETH criterion is satisfied in a slower rate than the rate of the Hilbert space expansion.

Turning to the second criteria, it is found that the maximum off-diagonal element of chaotic states tend to decrease with increasing atom number, see Figure~\ref{SMFigure3}c. Consistently with this observation, Figure~\ref{SMFigure3}d shows that the off-diagonal elements of the chaotic eigenstates smoothly vary with the energy differences where the red line is the moving average over the data (blue). This is a typical behavior for systems satisfying ETH \cite{Rigolreview}. The off-diagonal values of the regular states, in contrast, are spreading over all values within the machine precision, and do not vary smoothly as a  function of the energy differences, Figure~\ref{SMFigure3}e.

\section{Effective description of the regular states}

Next, we testify the self-consistency of the effective Hamiltonian, $\hat H_{\rm eff}$, description analyzed in the main text for the regular eigenstates. This Hamiltonian is obtained after neglecting the coupling $0 \leftrightarrow -1$, which we expect to hold in the $N_+\gg N_-$ regime leading to a bosonic amplification of the coupling $0\leftrightarrow 1$. 
As a next step, we discuss the role of finite size effects and  finally, based on the effective description, we calculate two approximate spectrum generating algebras which can be used to  construct the regular states.

\subsection{Effective description and self-consistency}

\begin{figure}
\centering
\includegraphics[width=0.7\columnwidth]{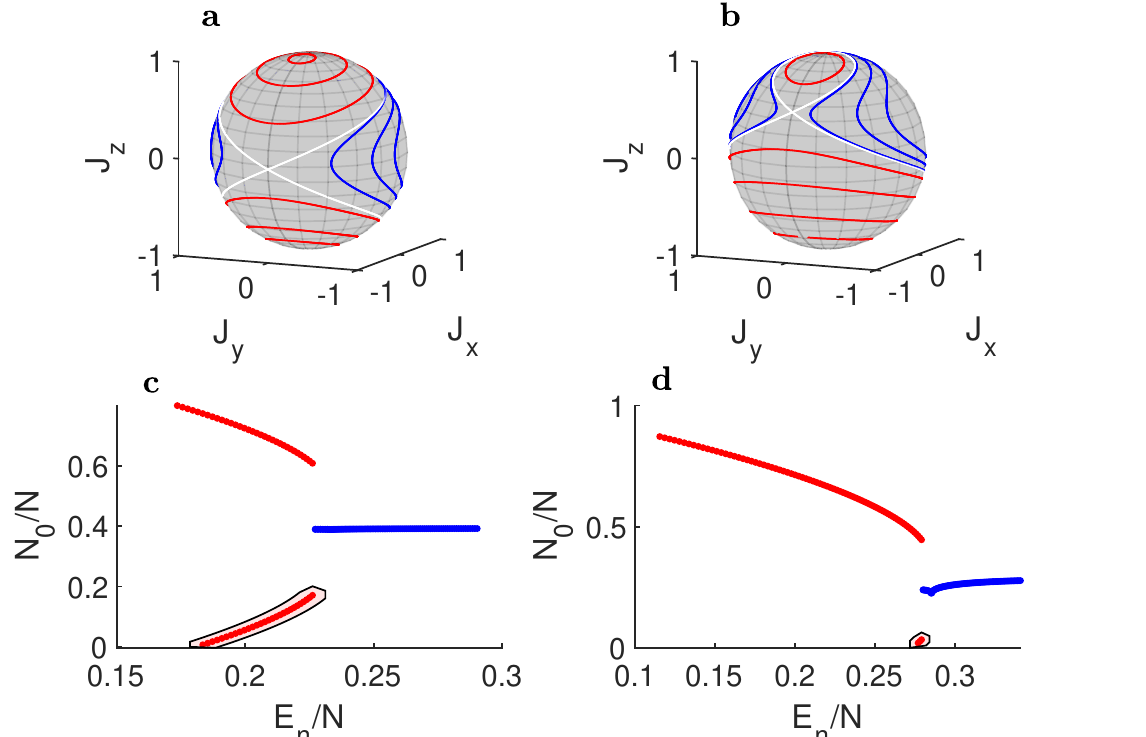}
\caption{{\bf a} and {\bf b}: Equal energy contours of the effective spin Hamiltonian of Eq.~\eqref{eq.Heff-supp} within the mean-field limit. The white lines represent the separatrix between the Fock regime (red lines) and the Rabi regime (blue lines). We consider $N=200$, $p_x=0.05$ and in {\bf a}, $N_-=39$ ($\delta \approx 0.013$) while in {\bf b}, $N_-=25$ ($\delta \approx 0.19$). 
In {\bf c} and {\bf d} Eigenstate expectation values for $\hat N_0/N = (J-\hat J_z)/N$. The eigenstates corresponding to the Fock (Rabi) regime are shown in red (blue). The lower Fock branch (highlighted) refers to the eigenstates that reproduce a tower of regular states.}\label{SMFigure1}
\end{figure}

The effective Hamiltonian can be seen either as a two-site Bose-Hubbard model, also known as the bosonic Josephson Junction \cite{gati2007bosonic} or equivalently can be viewed as a spin model for the collective spin degree of freedom $\hat J = \sum[\sigma_{\nu}]_{j,k}\hat a^\dagger_j\hat a_k$, where $\sigma_{\nu}$ are the Pauli matrices,  $\nu =x,y,z$ and $j,k\in\{0,+1\}$. In this latter representation it reads 
\begin{align}
    \hat H_{\rm eff}(N_-) = -\frac{\hat J_z^2}{2N}+\delta\hat J_z
    +\sqrt{2}p\hat J_x+C\,,\label{eq.Heff-supp}
\end{align}
where $\hat J_z = (\hat N_+-\hat N_0)/2$, $\delta = (N-5N_-)/(2N)$ and $C=-(3N^2-6NN_-+7N_-^2)/(8N)$. Due to the exchange symmetry, the system is in the spin representation $J=(N_++N_0)/2$ and a mean-field state can be fully parameterized by the two angles $\Theta$ and $\Phi$ in spherical coordinates. The mean-field energy is then obtained following the substitutions $\hat a_0\to (2J)^{1/2}\sin(\Theta/2)\exp(\ii\Phi)$ and $\hat a_+\to (2J)^{1/2}\cos(\Theta/2)$. We present the resulting equal energy contours for two different values of $N_-$ in Figure\,\ref{SMFigure1}\,{\bf a} and \,{\bf b} respectively. 
It becomes evident that two regimes can be distinguished based on the topology of the orbits on the unit sphere. 
In particular, starting from the poles there are  
orbits (see red lines) encircling the sphere, which correspond to the so-called Fock regime \cite{gati2007bosonic}. On the other hand, near the equator we have the Rabi regime (see blue lines), where $\phi$ is bounded on a given orbit. 
It should be clarified that in the Fock regime, the eigenstates are close to Fock states, by means that the mode populations exhibit sub-Poissonian fluctuations, while within the Rabi regime the eigenstates are closer to the eigenstates of the collective spin operator $\hat{J}$, and have large population fluctuations.

Having categorized the eigenstates, we turn back to our initial approximation in order to identify the impact of neglecting the mode coupling term $W_{-}$.
In other words, see also our above discussion, we aim to verify that the $m=-1$ mode is close to a Fock state. Hence, we neglect the $+1\leftrightarrow 0$ mixing (related to the term $\hat W_+$), to extract an effective Hamiltonian of the form of Eq.~\eqref{eq.Heff-supp} for the modes $m=-1,0$ where $N_+ \in \mathbb{Z}$. This approximation is certainly not justified in case that the $+1\leftrightarrow0$ mixing belongs to the Rabi regime, but it is sensible in the Fock regime. Let us now check whether the $-1\leftrightarrow0$ mixing term is also in the Fock regime for the self-consistency of the approach. This depends on the value of $N_+$ which determines the bias $\delta = (N-5N_+)/(2N)$. For sufficiently large $N_+$ (Fock regime around the north pole of the unit sphere in Figure\,\ref{SMFigure1}), we find that the effective Hamiltonian for the $0\leftrightarrow-1$ mixing is always deep in the Fock regime, and therefore our approach is self consistent. For smaller $N_+$ (Fock regime around the south pole), the $0\leftrightarrow-1$ mixing can access the Rabi regime, so that taking the $m=-1$ mode to be in the Fock state is not self-consistent. This explains why only the eigenstate of $\hat H_{\rm eff}$ of the Fock regime with $N_+>N_0$, which constitute the lower branch in Figure\,\ref{SMFigure1}\,{\bf c},{\bf d} have a corresponding tower of regular eigenstates in the three-mode spinor Hamiltonian. Upon tuning $N_-$, we change the values of the bias $\delta$ in Eq.~\eqref{eq.Heff-supp} and obtain different towers of regular eigenstates, until the point where the Fock regime on the north pole disappears (almost achieved for the parameters of Figure\,\ref{SMFigure1}\,{\bf b},{\bf d}).

We have made an arbitrary choice at the beginning of our argumentation when assuming $N_+>N_-$ in order to motivate that the mixing $+1\leftrightarrow0$ would dominate the dynamics. It is, of course, possible to equally choose $N_->N_+$, and reach the same conclusion. As mentioned in the main text, the regular states indeed come in pairs of states related by inversion symmetry.

\subsection{Detecting finite size effects}

\begin{figure*}
\centering
\includegraphics[]{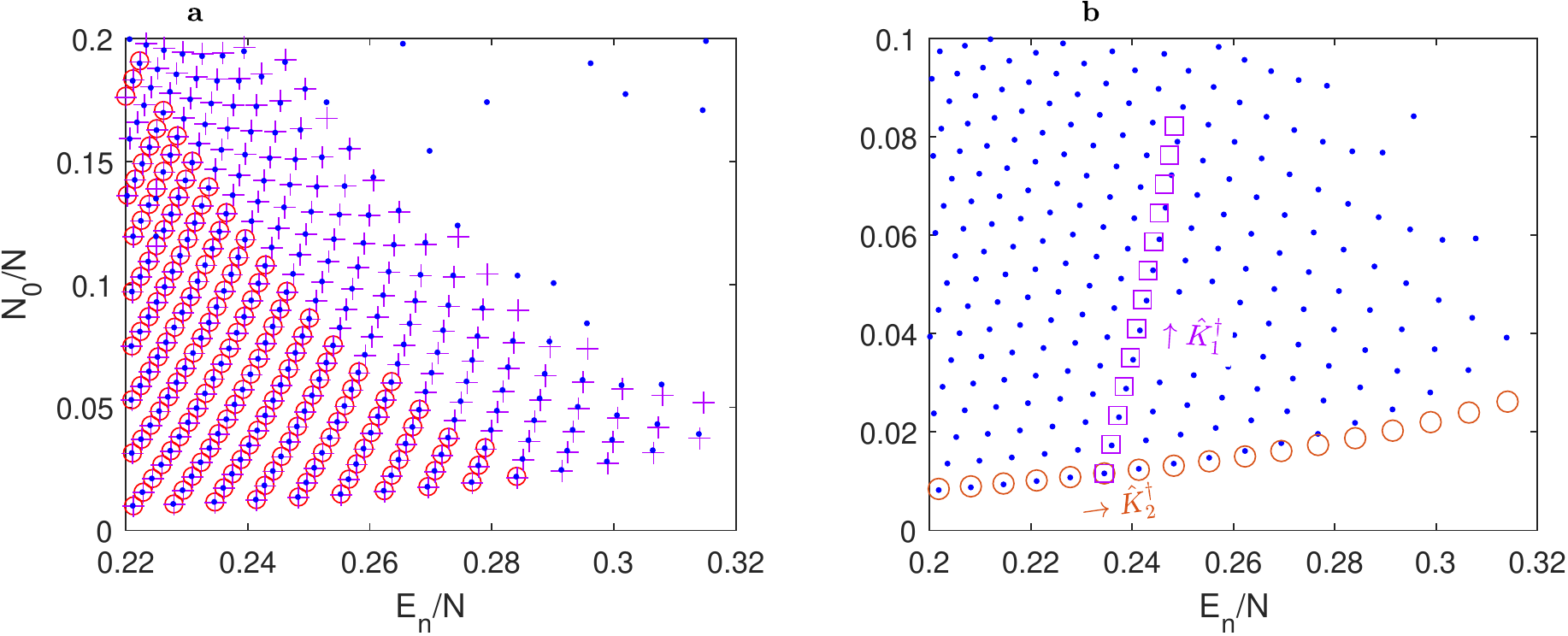}
\caption{Towers of regular states. The blue dots are obtained from exact diagonalization of the three-mode Hamiltonian for $N=200$. In {\bf a} the red circles refer to the eigenstates of the effective two-mode Hamiltonian $\hat H_{\rm eff}$ for $N=200$. The purple crosses are one out of ten eigenstates of $\hat H_{\rm eff}$ for $N=2000$. In {\bf b} shown are the states constructed by the repeated action of the operators $\hat K_2$ and $K_2^\dagger$ to jump in between towers (red circles) and the ones constructed by the application of $\hat K_1^\dagger$ to move up a tower. Our starting point is the exact ground state of the tower $N_-=31$ (inside both a square and a circle).}\label{SMFigure2}
\end{figure*}

Considering the atom number $N=200$ that has been mainly used in the main text, only about half of the regular eigenstates can be described by an eigenstate of our effective Hamiltonian of Eq.~(\ref{eq.Heff-supp}). 
We argue that this is a manifestation of finite size effects. 
To understand their role, we diagonalize $\hat H_{\rm eff}$ employing a much larger atom number, e.g.,~$N'=10N$ and we pick one eigenstate out of ten ($N'/N$). 
Following this procedure, it is possible to reproduce a full tower of regular states (providing that we study normalized quantities, such as $E_n/N$ or $\mean{\hat N_0}_n/N$). 
This can be readily seen in Figure\,\ref{SMFigure2}\,{\bf a} where the purple crosses coincide with the predictions of the three-mode Hamiltonian. 

\subsection{Approximate spectrum generating algebras} 

Quantum many-body scars can often be constructed by the  repeated action of an operator $\hat K^\dagger$ satisfying a so-called spectrum generating algebra $[\hat H,\hat K^\dagger]=\hat K^\dagger$ as it was discussed in Refs.~\cite{2021NatPh..17..675S,Turner2018,PhysRevB.102.085140,regnault2022quantum}. We can obtain the regular states of our model in a similar fashion, albeit it is here only an approximate construction. Our starting point is the effective Hamiltonian of Eq.~(\ref{eq.Heff-supp}). In the limit $N_0\ll N_+$, we linearize the effective Hamiltonian using $\hat J_z^2\approx 2J\hat J_z-J^2$ and obtain
\begin{align}
    \hat H_{\rm eff}(N_-)&\approx -\frac{2U}{N} N_-\hat J_z+\sqrt{2}p\hat J_x+\frac{UJ^2}{2N}+C\,
    \approx -\omega\left(\cos\alpha\hat J_z+\sin\alpha\hat J_x\right)+C\,,
\end{align}
where $\omega=[(2UN_-/N)^2+2p^2]^{1/2}$, $\alpha=-\arctan[pN/(\sqrt{2}UN_-)]$ and $C$ is a constant. Next, introducing the eigenmodes  
\begin{align}
    \hat b_+=\cos\left(\frac{\alpha}{2}\right)\hat a_++\sin\left(\frac{\alpha}{2}\right)\hat a_0\,,~~~\textrm{and}~~~
    \hat b_0=-\sin\left(\frac{\alpha}{2}\right)\hat a_++\cos\left(\frac{\alpha}{2}\right)\hat a_0\,,
\end{align}
we arrive to the following form 
\begin{align}
    \hat H_{\rm eff}\approx\frac{\omega}{2}\left(\hat b_0^\dagger\hat b_0-\hat b_+^\dagger\hat b_+\right)+C'\,.
\end{align}
Starting from the ground state $\ket{\psi_{N_-,0}}=\frac{\hat b_+^{\dagger (N-N_-)}}{\sqrt{(N-N_-)!}}\ket{0}$, we construct a tower of regular states $\ket{\psi_{N_-,n}}$ after the repeated application of the operator $\hat K_1=\hat b_0^\dagger\hat b_+$. 
To construct the ground state $\ket{\psi}_{N_-+1,0}$ of the adjacent tower with $N_-'=N_-+1$, we need to i) create an atom at $m=-1$, ii) reduce the pseudo-spin $J$ by $1/2$, and iii) rotate the spin by $\delta\alpha=\alpha(N_-+1)-\alpha(N_-)\approx 2\sqrt{2}p/(N\omega^2)\ll1$ in order to remain in the ground state (note that these operations commute to first order in $\delta\alpha$). This is achieved by the operator 
\begin{align}
    \hat K_2=[\hat N_-(N+1-\hat N_-)]^{-1/2}\hat a_-^\dagger\hat b_+\ee^{-\ii\delta\alpha\hat J_y}\,,
\end{align}
where the prefactor $[\hat N_-(N+1-\hat N_-)]^{-1/2}$ is needed for the normalization. Note that we have kept the exact expression of the rotation operator for clarity, but in the limit where this calculation is valid, we can use $\ee^{-\ii\delta\alpha\hat J_y}\approx1-\ii\delta\alpha\hat J_y$.

The eigenstates constructed using the operators $\hat K_1$ and $\hat K_2$ are presented in Figure \ref{SMFigure2}. 
As it can be seen, there is a rather good agreement for the states with $N_0/N\ll1$, where the linearization of the effective Hamiltonian is justified. In principle, we can construct a state of $J=(N_++N_0)/2$ by the application of $\hat K_1^\dagger$ for a fixed $N_-$, but the approximation of small $N_0/N\ll1$ eventually breaks down. 
For this reason, the states shown in Figure\,\ref{SMFigure2} correspond up to the 13 iteration, since at this point (chosen arbitrarily) the states that we construct deviate significantly from the exact eigenstates.
        
\section{Orbit stability analysis in the semiclassical limit}

A formal and general way of calculating the Lyapunov exponents is through constructing the so-called fundamental matrix, which is essential for the study of the stability of the trajectories. The main idea is to write down a set of coupled equations for the perturbations \cite{gaspard_1998, parker2012practical}. We have a set of equations that can be formally written,
\begin{eqnarray}
    \dot{\mathbf{x}} = F(\mathbf{x},t), \hspace{5mm} \mathbf{x}(t_0)=\mathbf{x}_0,\label{orbitSec:Eq:0}
\end{eqnarray}
$\mathbf{x}=[n_0,\theta,m,\eta]$, the set of parameters and $\mathbf{x}_0$ are the initial conditions. We can write a general solution as $\phi_t(\mathbf{x}_0,t_0)$ starting from the initial conditions $\mathbf{x}_0$. This can be substituted into Eq.~\eqref{orbitSec:Eq:0} to obtain,
\begin{eqnarray}
    \dot{\phi}_t(\mathbf{x}_0,t_0) = F(\phi_t(\mathbf{x}_0,t_0),t), \hspace{5mm} \phi_{t_0}(\mathbf{x}_0,t_0) = \mathbf{x}_0. \label{orbitSec:Eq:1}
\end{eqnarray}
Let us differentiate Eq.~\eqref{orbitSec:Eq:1} with respect to the initial condition $\mathbf{x}_0$,
\begin{eqnarray}
    D_{\mathbf{x}_0} \dot{\phi}_t(\mathbf{x}_0,t_0) = D_{\mathbf{x}} F(\phi_t(\mathbf{x}_0,t_0),t) D_{\mathbf{x}_0} \phi_t(\mathbf{x}_0,t_0), \hspace{5mm} D_{\mathbf{x}_0} \phi_{t_0}(\mathbf{x}_0,t_0) = \mathbb{I}_4. \label{orbitSec:Eq:2}
\end{eqnarray}
We define a new matrix-valued variable $\Phi_t(\mathbf{x}_0,t_0) := D_{\mathbf{x}_0} \phi_t(\mathbf{x}_0,t_0)$, and write Eq.~\eqref{orbitSec:Eq:2} as
\begin{eqnarray}
    \dot{\Phi}_t(\mathbf{x}_0,t_0) = D_{\mathbf{x}} F(\phi_t(\mathbf{x}_0,t_0),t) \Phi_t(\mathbf{x}_0,t_0), \hspace{5mm} \Phi_{t_0}(\mathbf{x}_0,t_0) = \mathbb{I}_4.\label{orbitSec:Eq:3}
\end{eqnarray}
Equation \eqref{orbitSec:Eq:3} is called the variational equation \cite{gaspard_1998, parker2012practical}. It is a matrix-valued time-varying linear differential equation. It originates from the linearization of the vector field along the trajectory $\phi_t(\mathbf{x}_0,t_0)$. Hence, the variational equation depends on the trajectory, which itself depends on the initial condition. A perturbation $\delta \mathbf{x}_0$ on the initial condition $\mathbf{x}_0$ evolves as
\begin{eqnarray}
    \delta \mathbf{x}(t) = \Phi_t(\mathbf{x}_0,t_0) \delta \mathbf{x}_0,
\end{eqnarray}
due to the initial condition of the variational equation being the identity matrix. $\Phi_t(\mathbf{x}_0,t_0) $ is called the fundamental matrix. Hence, to find the fundamental matrix, we need to solve the original set of equations together with the variational equation, namely
\begin{eqnarray}
    \dot{\mathbf{x}} &=& F(\mathbf{x},t),\notag \\
    \dot{\Phi} &=& D_{\mathbf{x}} F(\mathbf{x},t) \Phi, \label{orbitSec:Eq:4}
\end{eqnarray}
with the initial condition
\begin{eqnarray}
    \mathbf{x}(t_0) = \mathbf{x}_0, \hspace{5mm}\Phi(t_0) = \mathbb{I}_4.
\end{eqnarray}
The Lyapunov exponent associated with the initial state $\mathbf{x}_0$ can be computed via the fundamental matrix as,
\begin{eqnarray}
    \lambda(\mathbf{x}_0) = \lim_{t\rightarrow \infty} \frac{1}{t} \log || \Phi_t(\mathbf{x}_0,t_0) ||. \label{orbitSec:Eq:6}
\end{eqnarray}
Our equations of motion in $4D$ phase space are,
\begin{eqnarray}
\dot n_0 &=& p \sqrt{n_0}\left[\sqrt{1+m-n_0}\sin \left( \frac{\theta+\eta}{2} \right)+ \sqrt{1-m-n_0}\sin \left( \frac{\theta-\eta}{2} \right) \right]\notag \\
	\dot \theta &=& 2(1-2n_0) +p\left[\frac{1+m-2n_0}{\sqrt{n_0(1+m-n_0)}}\cos \left( \frac{\theta+\eta}{2} \right)  +\frac{1-m-2n_0}{\sqrt{n_0(1-m-n_0)}}\cos \left( \frac{\theta-\eta}{2} \right)  \right] \notag \\
	\dot m &=& p\sqrt{n_0}\left[-\sqrt{1+m-n_0}\sin  \left( \frac{\theta+\eta}{2} \right) +\sqrt{1-m-n_0}\sin \left( \frac{\theta-\eta}{2} \right) \right]\notag \\
	\dot\eta &=& -2m-2p_z-p \sqrt{n_0} \left[\frac{1}{\sqrt{1+m-n_0}}\cos\left( \frac{\theta+\eta}{2} \right) -\frac{1}{\sqrt{1-m-n_0}}\cos\left( \frac{\theta-\eta}{2} \right) \right].\label{orbitSec:Eq:5}
\end{eqnarray}
We can express the set of differential equations for the perturbations as,
\begin{eqnarray}
    \delta \dot{\mathbf{x}}(t) = \frac{\delta F(\mathbf{x},t)}{\delta n_0} \delta n_0 + \frac{\delta F(\mathbf{x},t)}{\delta \theta} \delta \theta + \frac{\delta F(\mathbf{x},t)}{\delta m} \delta m + \frac{\delta F(\mathbf{x},t)}{\delta \eta} \delta \eta  
\end{eqnarray}
These derivatives can be exactly calculated but since they do not provide additional insights, we omit writing them here. By introducing the abbreviation 
for the derivatives $\delta_{x_i}F_{x_j} \equiv d F_j(\mathbf{x},t)/d x_i$ the matrix $D_{\mathbf{x}} F(\mathbf{x},t)$ takes the form 
\begin{eqnarray}
    D_{\mathbf{x}} F(\mathbf{x},t) &=& \left(\begin{array}{cccc}
        \delta_{n_0}F_{n_0} & \delta_{\theta}F_{n_0} & \delta_{m}F_{n_0} & \delta_{\eta}F_{n_0} \\
        \delta_{n_0}F_{\theta} & \delta_{\theta}F_{\theta} & \delta_{m}F_{\theta} & \delta_{\eta}F_{\theta} \\   
        \delta_{n_0}F_{m} & \delta_{\theta}F_{m} & \delta_{m}F_{m} & \delta_{\eta}F_{m} \\  
        \delta_{n_0}F_{\eta} & \delta_{\theta}F_{\eta} & \delta_{m}F_{\eta} & \delta_{\eta}F_{\eta} 
    \end{array}\right).
\end{eqnarray}
Note that $ \delta_{\theta}F_{\eta} =  \delta_{m}F_{n_0}$, $\delta_{\eta}F_{\eta} = - \delta_{m}F_{m}$, $\delta_{\eta}F_m = -\delta_{\theta}F_{n_0}$, $\delta_{\theta}F_m = - \delta_{\eta}F_{n_0}$, $\delta_{n_0}F_{n_0} = -\delta_{\theta}F_{\theta}$ and $\delta_{\eta}F_{\theta} = \delta_{n_0} F_m$ based on the spinor condensate mean-field equations. Then we can numerically solve Eq.~\eqref{orbitSec:Eq:4} to find the Lyapunov exponents.

Numerical integration of the variational Eq.~\eqref{orbitSec:Eq:4} can be done for any finite time when the trajectory is stable approaching $\lambda(\mathbf{x_0},t) \rightarrow 0$ in the infinite time limit $t\rightarrow \infty$. In contrast, when the trajectory is chaotic, numerical error accumulating eventually diverges resulting in a limited simulation time, albeit a long one which exhibits an equilibration in $\lambda(\mathbf{x_0},t)$. We average $\lambda(\mathbf{x_0},t)$ over the accessible time interval, and determine the Lyapunov exponents $\lambda(\mathbf{x}_0) \sim 1/T \int_0^{T} dt\hspace{1mm} \lambda(\mathbf{x_0},t)$. This is the analysis performed to find the maximum positive Lyapunov exponent out of four in a 4D phase space as reported in Figure 2a of the main text. 

Let us examine the simplest case analytically here. Note that there is a fixed point in the equations of motion Eq.~\eqref{orbitSec:Eq:5} at $\mathbf{x}_0=[0.5,0,0,0]$. At this fixed point, we have the Jacobian matrix
\begin{equation}
D_{\mathbf{x}} F(\mathbf{x},t) =
\begin{pmatrix}
        0 & p/2 & 0 & 0\\
         -4(1+2p) & 0 & 0 & 0\\
       0 & 0 & 0 & -p/2\\
        0 & 0 & -2+2p & 0
\end{pmatrix}.
\end{equation}
Since this is a fixed point, at $t\rightarrow \infty$, we still have the same expression. Let us calculate the eigenvalues,
\begin{equation}
\begin{aligned}
    & \sigma_{1,2} = \pm \sqrt{-2p(1+2p)} \\
    & \sigma_{3,4} = \pm \sqrt{p - p^2},
\end{aligned}
\end{equation}
with the corresponding eigenstates
\begin{equation}
    v_{1,2} \propto \begin{pmatrix}
        \mp\frac{1}{2}\sqrt{\frac{-p}{2(1+2p)}} \\
        1 \\
        0 \\
        0
    \end{pmatrix},
    \quad
    v_{3,4} \propto \begin{pmatrix}
        0 \\
        0 \\
        \pm\frac{p}{2}\frac{1}{\sqrt{p-p^2}}  \\
        1
    \end{pmatrix}.
\end{equation}
Depending on $p$, each pair of eigenvalues will be either a pair of opposite real numbers, or a pair of complex conjugate numbers. For the region of interest, i.e. $p=0.05$, we examine the Lyapunov exponents for $0 < p < 1$ where we have $\sigma_{1,2} \in \mathbb{C}$. These correspond to circles (periodic orbits) on the plane $(n_0,\theta)$ defined by the respective eigenvectors, $v_{1,2} \in \mathbb{C}$. These orbits will be followed with a frequency $\omega = |\sigma_{1,2}| = \sqrt{2p(1 + 2p)}$. This is consistent with the family of unstable periodic orbits (UPO) identified on the plane $(n_0,\theta)$, and discussed in detail in the main text. 
On the other hand, $\sigma_{3,4} \in \mathbb{R}$ associated with $v_{3,4} \in \mathbb{R}$, correspond to a saddle point of the dynamics. Out of plane perturbations will therefore deviate from the fixed point exponentially, with the Lyapunov exponent being equal to $\lambda = \sqrt{p-p^2}$. Let us note that we could also study the associated fundamental matrix $\Phi_t(\mathbf{x}_0,t_0)$ at this fixed point. At any time, the eigenvalues of the upper block of the fundamental matrix, i.e.,~the phase space related to the periodic orbit on the plane, are $1$ resulting in $\lambda=0$ Lyapunov exponents for the in-plane perturbations. Hence, the UPO is stable to in-plane perturbations. On the other hand, the eigenvalues of the lower block of the fundamental matrix are a function of time, such that the Lyapunov exponent Eq.~\eqref{orbitSec:Eq:6} results in $\lambda = \sqrt{p-p^2}$.

\begin{figure}
\centering
\includegraphics[]{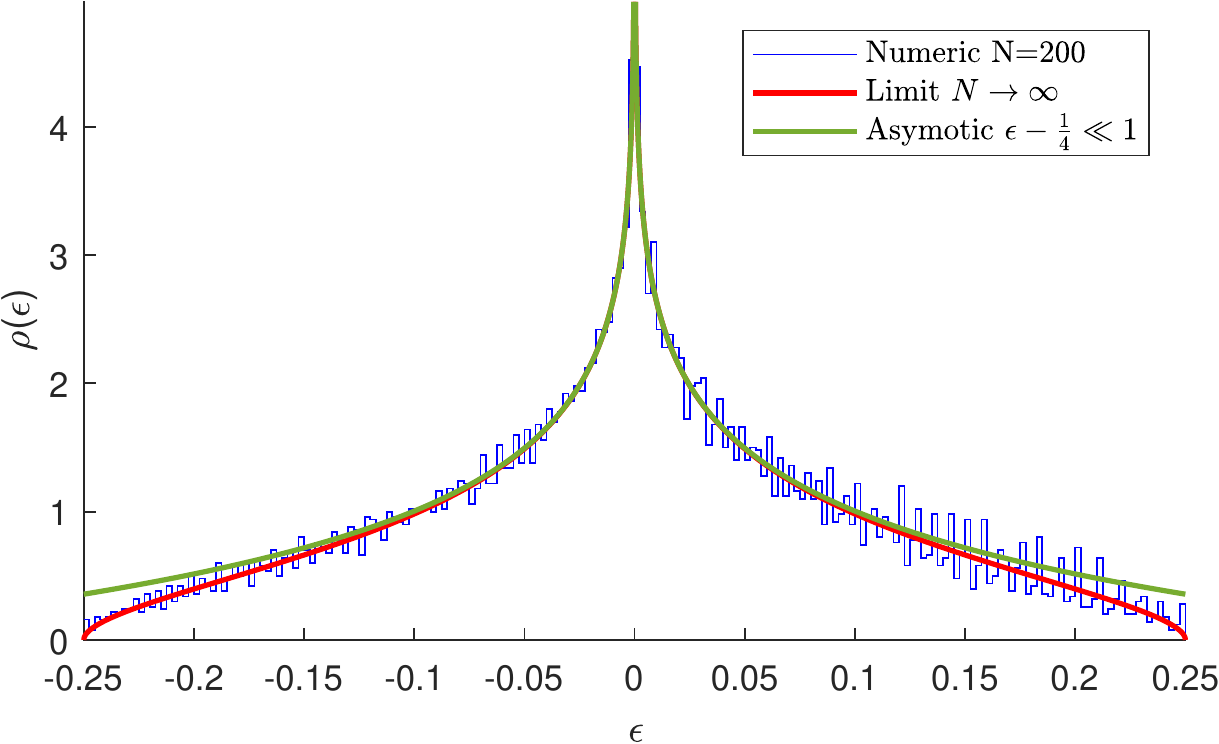}
\caption{Density of states of the interaction Hamiltonian, obtained numerically for $N=200$ (blue line) and compared to the expressions of Eq.~(\ref{eq.DOS}) (red line) and Eq.~(\ref{eq.DOSapprox}) (green line).}\label{SMFigureDOS}
\end{figure}

\section{Density of states of the interaction Hamiltonian}

An important feature of our interaction Hamiltonian $\hat H_{\rm int}$ is a divergence of the density of states in the middle of the energy spectrum, which allows for a rapid emergence of chaotic behavior. 
Notice that this divergence is absent in the standard Hamiltonian of the spinor condensate with spin-changing collision terms (see above), and it turns out to be crucial to observe the coexistence of the regular and thermal states. Below, we briefly sketch the calculation of the density of states. 

The eigenstates of the interaction Hamiltonian are the Fock states $\ket{N_+,N_0,N_-}$ where $N_m$ is the eigenvalue of the operator $\hat N_m$. 
To simplify the derivation we adopt the quantum numbers $N_0$, $M=N_+-N_-$ and $N=N_++N_0+N_-$. Then, the energy of an eigenstate of $\hat H_{\rm int}$ (assuming $c_1=1$ in the following) reads
\begin{align}
    E_{\rm int}(N_0,M,N)=\frac{1}{N}\left[N_0\left(N-N_0\right)+\frac{M^2}{2}\right]\,.
\end{align}
Accordingly, the density of states at energy $E$, for a fixed atom number $N$ takes the form 
\begin{align}
\rho(E)=\frac{2}{W}\sum_{N_0=1}^{N}\sum_{M=0}^{1-N_0}\delta\left[E-E_{\rm int}(N_0,M,N)\right]\,.
\end{align}
The normalization constant, $W=N/2$, refers to the energy bandwidth and the factor $2$ stems from the fact that we restrict the sum over $M$ to only positive values, taking advantage of the inversion symmetry. 
Furthermore, considering the limit $N\to+\infty$, the density of states in the continuum limit reads
\begin{align}
    \rho(\epsilon)=4\int_0^{1}dn_0\int_0^{1-n_0}dm \delta \left[\epsilon-n_0(1-n_0)-\frac{m^2}{2}\right] = 2\sqrt{2}\left[f\left(\frac{x_2}{x_1}\right)-f\left(\frac{1}{x_1}\right)\right]\,,\label{eq.DOS}
\end{align}
where $\epsilon=E/N$, $n_0=N_0/N$, $m=M/N$, $x_1 = \sqrt{|4\epsilon-1|}$, $x_2=-1+2\sqrt{1-2\epsilon}$, $f=$acosh if $\epsilon<1/4$ and $f=$asinh if $\epsilon>1/4$. 
The above expression can be expanded around the middle of the energy spectrum, $\epsilon=1/4$, leading to  
\begin{align}
    \rho(\epsilon)\sim2\sqrt{2}\log\left(\left|\epsilon-\frac{1}{4}\right|\right)+2\sqrt{2}\log(\sqrt{2}-1)\,.\label{eq.DOSapprox}
\end{align}
It becomes evident that it shows a logarithmic divergence. 
A direct comparison of these two expressions with the numerically obtained density of states for $N=200$ is provided in Figure\,\ref{SMFigureDOS} where an excellent agreement is observed. 

\begin{figure}
\centering
\includegraphics[]{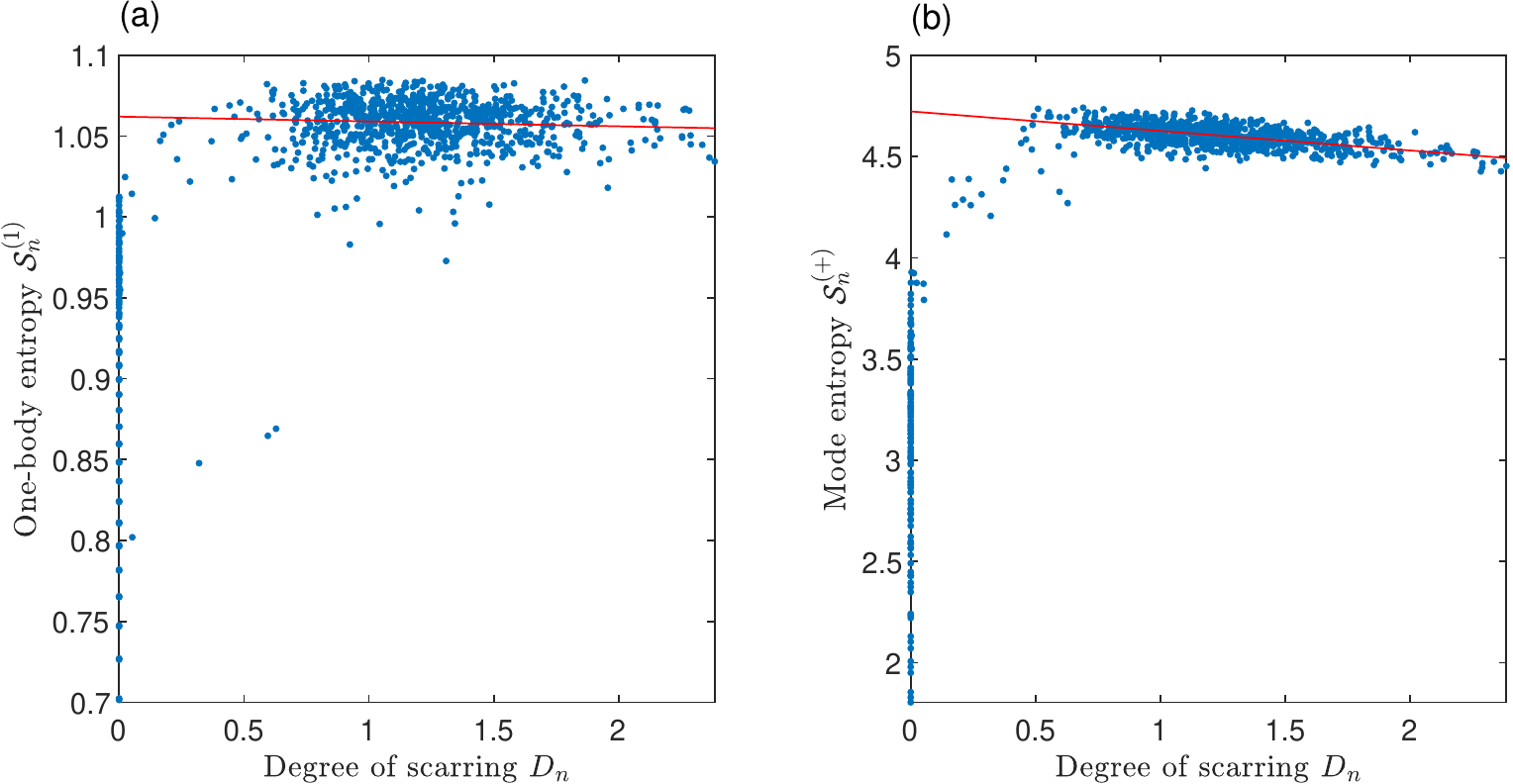}
\caption{(a) One-body and (b) mode entropy as a function of the scarring degree $D_n$. 
The red line represents a linear fit (excluding the regular state with $D_n\approx0$).}\label{SMFigureEntropies}
\end{figure}
\section{Entanglement entropy and scarring}

To further address the relation between entanglement and scarred eigenstates we relied on additional entropy measures than the one described in the main text. 
Recall that employing the von-Neumann entropy $\mathcal{S}^{(1)}_n=-\mathrm{Tr}(\rho^{(1)}_{n}\ln\rho^{(1)}_{n})$ of the one-body density matrix $[\rho^{(1)}_{n}]_{jk}=\bra{\psi_n}\hat a_j^\dagger\hat a_k\ket{\psi_n}$, it was demonstrated that scarred states are characterized by close-to-maximal entanglement entropy, in sharp contrast with regular eigenstates. Here, we additionally present the entropy obtained after a mode partitioning of the system.  Namely, the entropy $\mathcal{S}^{(+)}_n=-\mathrm{Tr}(\rho^{(+)}_{n}\ln\rho^{(+)}_{n})$ of the reduced density matrix for the $m=+1$ mode is evaluated after tracing over the other two-modes ($m=-1,0$) where  $\rho^{(+)}_{n}=\mathrm{Tr}_{-1,0}(\ket{\psi_n}\bra{\psi_n})$. 
Figure\,\ref{SMFigureEntropies} depicts how the above-mentioned  entanglement entropies behave with the degree of scarring $D_n$ introduced in the main text. 
Their overall trend is the same, i.e. there is a group of regular states with $D_n\approx0$ and relatively low entropy, and importantly a branch of scarred states, with $D_n\gtrsim1$ and high entropy. However, we note that while $S^{(1)}_n$ appears flat as a function of $D_n$, the mode entropy $S^{(+)}_n$ displays a slightly negative correlation tendency.

\end{document}